\documentclass{article}
\usepackage{amsmath}
\usepackage{amsfonts}
\usepackage{amssymb}
\usepackage{hyperref}

\begin{document}

\title{\textbf{Lie-deformed quantum Minkowski spaces from twists:
Hopf-algebraic versus Hopf-algebroid approach }}
\author{\textbf{Jerzy Lukierski}$^{1}$, \textbf{\ Daniel Meljanac}$^{2}$, 
\textbf{Stjepan Meljanac}$^{3}$, \\
\textbf{Danijel Pikuti\'{c}}$^{3}$, \textbf{Mariusz Woronowicz}$^{1}$ 
\vspace{12pt} \\
$^1$ Institute for Theoretical Physics, University of Wroclaw,\\
pl. Maxa Borna 9, 50-205 Wroclaw, Poland \\
$^2$Division of Materials Physics, Rudjer Bo\v{s}kovi\'c Institute, \\
Bijeni\v{c}ka c. 54, HR-10002 Zagreb, Croatia \\
$^3$Division of Theoretical Physics, Rudjer Bo\v{s}kovi\'c Institute, \\
Bijeni\v{c}ka c. 54, HR-10002 Zagreb, Croatia}
\maketitle

\begin{abstract}
We consider new Abelian twists of Poincare algebra describing nonsymmetric
generalization of the ones given in \cite{lw06}, which lead to the class of
Lie-deformed quantum Minkowski spaces. We apply corresponding twist
quantization in two ways: as generating quantum Poincare-Hopf algebra
providing quantum Poincare symmetries, and by considering the quantization
which provides Hopf algebroid describing class of quantum relativistic phase
spaces with built-in quantum Poincare covariance. If we assume that Lorentz
generators are orbital i.e.do not describe spin degrees of freedom, one can
embed the considered generalized phase spaces into the ones describing the
quantum-deformed Heisenberg algebras.
\end{abstract}

\section{Introduction}

Due to quantum gravity (see \cite{mead}-\cite{cikg}) as well as quantized
strings effects (see e.g. \cite{siebwitt,boer}) at Planck distances the
notion of classical space-time can not be maintained. The quantum-mechanical
space-time localization in the presence of gravitational interactions are
constrained by new type of bounds, extending Heisenberg uncertainty
relations to the measurements of pairs of space-time coordinates (see e.g. 
\cite{dfr-2})\footnote{%
This new type of uncertainty relations will be further called
Doplicher-Fredenhagen-Roberts (DFR) uncertainty relations.}. Algebraically
DSR uncertainty relations can be derived from the noncommutative structure
of quantum Minkowski space $\widehat{\mathcal{M}}(\widehat{x}_{\mu }\in 
\widehat{\mathcal{M}})$ with nonvanishing commutator $[\widehat{x}_{\mu },%
\widehat{x}_{\mu }]_{|\mu \neq \nu }$ proportional to $\lambda _{pl}^{2}$\
(Planck length $\lambda _{pl}\simeq 10^{-33}cm$).

The noncommutative structures linked with quantum-deformed dynamical
theories (e.g. quantum gravity) appeared recently in two-fold way:

\begin{enumerate}
\item[i.)] as quantum generalization of Lie-algebraic symmetries, described
by noncommutative Hopf algebras \cite{drinfe},\cite{majid},

\item[ii.)] as deformed quantum phase spaces, with modified deformed
canonical Heisenberg relations, described in the formalism of noncommutative
geometry by a generalization of Hopf algebras, called Hopf algebroids \cite%
{lu}-\cite{bohm}.
\end{enumerate}

Both noncommutative structures can be generated by twist quantization
procedure. If twist $\mathcal{F}$ of classical Poincare-Hopf algebra is
generated by classical $r$-matrix with the terms $P\wedge M$, where $P$
describes the fourmomenta and $M$ the Lorentz generators, the
noncommutativity of quantum space-time takes the Lie-algebraic form (see
e.g. \cite{lw06}). In this paper we shall consider new class of such twists
and present explicitly twisted quantum Poincare symmetries as well as Hopf
algebroid structure of corresponding quantum phase spaces. We add that Hopf
algebroid structures of quantum phase spaces with Lie-algebraic space-time
sector were already discussed (see e.g. \cite{mss}-\cite{meljsko}), however
mostly either without considering the twist quantizations \cite{mss} or with
twisted Hopf algebroids considered not explicitly but as a part of general
mathematical framework \cite{xu},\cite{bp}; see however \cite{meljsko}.

The classical Minkowski space $\mathcal{M}(x_{\mu }\in \mathcal{M})\footnote{%
Further we shall consider physical $D=4$ case, i.e. $\mu =0,1,2,3$.}$ is
fully determined if it is given as the irreducible four-vector
representation of Lorentz algebra $\mathcal{O}(3,1)$, with the action of
Lorentz generators $M_{\mu \nu }$ given by the adjoint action describing the
semidirect product $\mathcal{O}(3,1)\rtimes \mathcal{M}$%
\begin{equation}
M_{\mu \nu }\rhd x_{\rho }\equiv \lbrack M_{\mu \nu },x_{\rho }]=\eta _{\nu
\rho }x_{\mu }-\eta _{\mu \rho }x_{\nu }.  \label{act1}
\end{equation}%
One can supplement as well the relation%
\begin{equation}
M_{\mu \nu }\rhd p_{\rho }\equiv \lbrack M_{\mu \nu },p_{\rho }]=\eta _{\nu
\rho }p_{\mu }-\eta _{\mu \rho }p_{\nu },  \label{act2}
\end{equation}%
by observing that Poincare algebra with classical generators $g=(M_{\mu \nu
},p_{\mu })$ is also endowed with a semidirect product structure $\mathcal{O}%
(3,1)\rtimes \mathcal{T}$ $(p_{\mu }\in \mathcal{T}).$The relations (\ref%
{act1}-\ref{act2}) can be extended further by the formula\footnote{%
The relation (\ref{act3}) can be linked to Hopf-algebraic scheme (see e.g. 
\cite{majid}) if we consider $\mathcal{P}$ and $\mathcal{T}$ as dual
bialgebras, in classical case with primitive coproducts.}%
\begin{equation}
p_{\mu }\rhd x_{\rho }=-i\eta _{\mu \rho }.  \label{act3}
\end{equation}%
In such a way we obtain consistent classical action of Poincare algebra $%
\mathcal{P}$ on the Minkowski space $\mathcal{M}$, describing the cross
product $\mathcal{P}\#\mathcal{M}$\footnote{%
The semidirect product of two Lie algebras is again a Lie algebra, what is
generalized by the notion of smash product, which describes the algebraic
structure on the vector space $H\oplus V$, where $H=(M_{H},\Delta
_{H},\epsilon _{H},1_{H})$ is (unital and counital) bialgebra (in particular
Lie bialgebra) and $V$ is a unital $H$-module algebra, which may be
noncommutative. Cross-product algebra can be endowed with Hopf algebroid
structure \cite{lu},\cite{brzez}.} .

Important class of quantum Poincare algebras are described by twist
quantizations of classical Poincare algebra, with all deformation located
only in coalgebraic sector of Poincare-Hopf algebra $\mathbb{H}.$ In such a
case the twist $\mathcal{F}\equiv \mathcal{F}^{(1)}\mathcal{\otimes F}%
^{(2)}\in \mathcal{U}(g)\otimes \mathcal{U}(g)$ depends on classical
Poincare generators, and the formula for coproducts%
\begin{equation}
\Delta _{\mathcal{F}}(g)=\mathcal{F}\Delta _{0}(g)\mathcal{F}^{-1},
\end{equation}%
can be calculated using only the classical Poincare algebra commutators.

The quantum Minkowski space $\widehat{\mathcal{M}}(\widehat{x}_{\mu }\in 
\widehat{\mathcal{M}})$ in the case of twist quantization is again fully
specified if it is given as the irreducible four-dimensional module of the
respective twisted Poincare-Hopf algebra $\mathbb{H}_{\mathcal{F}}$%
\begin{equation}
\mathbb{H}_{\mathcal{F}}=(\mathcal{U}(\mathcal{P}),m,\Delta _{\mathcal{F}%
},S_{\mathcal{F}},\epsilon ),  \label{hmod}
\end{equation}%
where $S_{\mathcal{F}}$\ denotes twisted antipode (coinverse) and $\epsilon $
is a counit.

The standard Hopf-algebraic way of introducing the quantum Minkowski
coordinates $\widehat{x}_{\mu }$ is to consider quantum Poincare group and
use the Hopf-algebraic duality between the coproducts of $p_{\mu }$ and the
coordinates $\widehat{x}_{\mu }$ and introduce the notion of Heisenberg
double \cite{majid,luknow}. However, the twist $\mathcal{F}$ provides
directly the formula linking $x_{\mu }\in \mathcal{M}$ with $\widehat{x}%
_{\mu }\in \widehat{\mathcal{M}}$, by means of the relation which is a
special case of derived in Sect.2 star-product relation (\ref{fdefor}) (for $%
f(x)=x$)%
\begin{equation}
\widehat{x}_{\mu }=\left[ \left( \mathcal{F}^{-1}\right)
^{(1)}\triangleright x_{\mu }\right] \left( \mathcal{F}^{-1}\right) ^{(2)},
\label{xTwist}
\end{equation}%
with the action $\triangleright $ in (\ref{xTwist}) provided by the formulae
(\ref{act1}), (\ref{act3}). In such a way one can express the noncommutative 
$\widehat{x}_{\mu }$ in terms of classical phase space coordinates $(x_{\mu
},p_{\mu })$ and generators $M_{\mu \nu }$.

In this paper we plan to consider the generalization of twist considered in 
\cite{lw06}, with arbitrary symmetry of $\log \mathcal{F}$ as tensor
product. In such a way we introduce additional real parameter $u$ which for $%
u=\frac{1}{2}$ leads to antisymmetric $\log \mathcal{F}$ tensor and
antisymmetric classical $r$-matrix (this was the case considered in \cite%
{lw06} and \cite{bala}); other cases $u=0$ and $u=1$ correspond to maximally
nonsymmetric twists, which for $u=0$ was discussed as well in the literature 
\cite{kumar}.

The aim of our paper is to describe the twist quantizations in two
frameworks: first is based entirely on Hopf-algebraic techniques, which
provides quantum Poincare-Hopf algebra $\mathbb{H}_{\mathcal{F}}$ and
second, which leads for any value of parameter $u$ to the embedding of $%
\mathbb{H}_{\mathcal{F}}$ into the Hopf algebroids describing suitably
deformed smash product $\mathcal{T}\#\mathcal{M}$. New results\ in the
present paper are provided by the second method by providing the Hopf
algebroid structure: construction of source, target and antipode maps (for
their definition see \cite{xu},\cite{bp}) and by determining the coproduct
freedom for twisted bialgebroids $\mathcal{H}_{\mathcal{F}}=(\mathcal{P}\#%
\mathcal{M)}_{\mathcal{F}}$ (so-called coproduct gauges, see \cite{lws}). We
add that the Hopf algebroid techniques were extensively studied in
mathematics (see e.g. \cite{takeu},\cite{lu}-\cite{bp},\cite{schen}) and
applied to the description of quantum-deformed relativistic phase spaces by
some of the present authors \cite{pachol-strajn}-\cite{jkm},\cite{lws}. The
novelty of our discussion of Hopf algebroid structure in comparison with our
earlier efforts \cite{pachol-strajn}-\cite{jkm}\ is to consider Lorentz
generators $M_{\mu \nu }$\ as independent - we shall not assume the standard
orbital phase space realization of $M_{\mu \nu }$%
\begin{equation}
M_{\mu \nu }=i(x_{\mu }p_{\nu }-x_{\nu }p_{\mu }).  \label{repM}
\end{equation}%
If relation (\ref{repM}) is valid, the Hopf-algebraic Poincare algebra twist 
$\mathcal{F}$ as well as noncommutative Minkowski space coordinates $%
\widehat{x}_{\mu }$ can be expressed in terms of phase space variables i.e.
the Hopf-algebraic formulae are realized in terms of canonical Heisenberg
algebra, which provides a classical example of Hopf algebroid. The
noncommutative Minkowski coordinates $\widehat{x}_{\mu }$ can be expressed
as the following functions of classical phase space variables $(x_{\mu
},p_{\mu })$%
\begin{equation}
\widehat{x}_{\mu }=x_{\nu }\varphi _{\ \mu }^{\nu }(p).  \label{xdef}
\end{equation}%
We mention that the formula (\ref{xdef}) was considered \cite{kumar}, \cite%
{ex1}-\cite{ex2} as well for other Lie-algebraic quantum deformations of
Poincare algebra, not necessarily described by twist quantization\footnote{%
Such example of quantization which can not be obtained by using twist is the 
$\kappa $-deformation of Poincare algebra \cite{luru-1}-\cite{luru-2}.} (see 
\cite{jkm,mss}).

\section{Twist-deformed Poincar\'{e} Hopf algebra and quantum Minkowski
spaces}

Poincar\'{e} algebra $\mathcal{P}$, generated by Lorentz generators $M_{\mu
\nu }$ and momentum generators $p_{\mu }$ is defined by 
\begin{align}
\lbrack p_{\mu },p_{\nu }]& =0,  \label{pkom} \\
\lbrack M_{\mu \nu },p_{\rho }]& =\eta _{\nu \rho }p_{\mu }-\eta _{\mu \rho
}p_{\nu }, \\
\lbrack M_{\mu \nu },M_{\rho \sigma }]& =\eta _{\nu \rho }M_{\mu \tau }-\eta
_{\mu \rho }M_{\nu \tau }-\eta _{\nu \tau }M_{\mu \rho }+\eta _{\mu \tau
}M_{\nu \rho },
\end{align}%
where $\eta _{\mu \nu }=diag(-1,1,...,1)$.

Classical Poincar\'{e} Hopf algebra is defined by the universal enveloping
algebra $\mathcal{U}(\mathcal{P})$ of the Poincar\'{e} algebra $\mathcal{P}$%
, together with the coproduct $\Delta _{0}$, antipode $S_{0}$ and counit $%
\epsilon _{0}$, given by 
\begin{align}
\Delta _{0}(p_{\mu })=p_{\mu }\otimes 1+1\otimes p_{\mu }& ,\qquad \Delta
_{0}(M_{\mu \nu })=M_{\mu \nu }\otimes 1+1\otimes M_{\mu \nu },
\label{undef-antip} \\
S_{0}(p_{\mu })=-p_{\mu }& ,\qquad S_{0}(M_{\mu \nu })=-M_{\mu \nu },
\label{anti} \\
\epsilon _{0}(p_{\mu })=0& ,\qquad \epsilon _{0}(M_{\mu \nu })=0.
\end{align}

Twist $\mathcal{F}$ is an invertible element of $\mathcal{U}(\mathcal{P}%
)\otimes \mathcal{U}(\mathcal{P})$, satisfying the cocycle condition 
\begin{equation}
(\mathcal{F}\otimes 1)(\Delta _{0}\otimes 1)\mathcal{F}=(1\otimes \mathcal{F}%
)(1\otimes \Delta _{0})\mathcal{F},  \label{cocycle-c}
\end{equation}%
and the normalization condition 
\begin{equation}
(\epsilon \otimes 1)\mathcal{F}=(1\otimes \epsilon )\mathcal{F}=1\otimes 1.
\label{normalization-c}
\end{equation}

Twists deform coproducts (\ref{undef-antip}) and antipodes (\ref{anti}) of $%
h\in \mathcal{U}(\mathcal{P})$ as follows: 
\begin{align}
\Delta _{\mathcal{F}}h& =\mathcal{F}\Delta _{0}h\mathcal{F}^{-1},
\label{general-def-copro} \\
S_{\mathcal{F}}(h)& =\chi _{\mathcal{F}}S_{0}(h)\chi _{\mathcal{F}}^{-1},
\label{general-def-antip}
\end{align}%
where $\chi _{\mathcal{F}}=m[(S_{0}\otimes 1)\mathcal{F}]$ and the deformed
coproduct $\Delta _{\mathcal{F}}$ is coassociative due to the cocycle
condition \eqref{cocycle-c}.

Here we consider the following families of Abelian twists%
, for all dimensions $n\geq 3$: 
\begin{equation}
\begin{split}
\mathcal{F}_{u}& =\exp \left( (1-u)\frac{a\cdot p}{2\kappa }\otimes \theta
^{\alpha \beta }M_{\alpha \beta }-u\theta ^{\alpha \beta }M_{\alpha \beta
}\otimes \frac{a\cdot p}{2\kappa }\right) \\
& =\exp \left( \frac{1-u}{2}\mathcal{K}^{\alpha \beta }\otimes M_{\alpha
\beta }-\frac{u}{2}M_{\alpha \beta }\otimes \mathcal{K}^{\alpha \beta
}\right) ,
\end{split}
\label{twist-F}
\end{equation}%
where $a\cdot p=a^{\mu }\eta _{\mu \nu }p^{\nu }$.

The parameter $u\in \lbrack 0,1]$, $\kappa $ is the deformation parameter
with dimension of mass, $a^{2}\in \{-1,0,1\}$, 
\begin{equation}
\mathcal{K}^{\mu \nu }=\frac{a\cdot p}{\kappa }\theta ^{\mu \nu },
\label{mathcalK}
\end{equation}%
and the following conditions hold: 
\begin{equation}
\theta ^{\mu \nu }=-\theta ^{\nu \mu },\qquad a_{\mu }\theta ^{\mu \nu }=0.
\end{equation}%
These twists are a generalization of the twist proposed in \cite{lw06} -
being Abelian, they automatically satisfy the cocycle condition. Because
under flip transformation $(a\otimes b)^{\tau }=b\otimes a$, we obtain the
following $u$\textit{-independent }formula for universal $\mathcal{R}$%
-matrix $(a\wedge b=a\otimes b-b\otimes a)$%
\begin{equation}
\mathcal{R}=\mathcal{F}_{u}^{\tau }\mathcal{F}_{u}^{-1}=\exp [\frac{1}{2}%
(M_{\alpha \beta }\wedge \mathcal{K}^{\alpha \beta })].
\end{equation}

Using equation \eqref{general-def-copro} for the family of twists %
\eqref{twist-F} we obtain the following deformed coproducts $(\mathcal{F}%
\equiv \mathcal{F}_{u})$%
\begin{align}
\Delta _{\mathcal{F}}(p_{\mu })& =p_{\alpha }\otimes (e^{-u\mathcal{K}%
})^{\alpha }{}_{\mu }+(e^{(1-u)\mathcal{K}})^{\alpha }{}_{\mu }\otimes
p_{\alpha },  \label{Delta-p} \\
\Delta _{\mathcal{F}}(M_{\mu \nu })& =M_{\alpha \beta }\otimes (e^{-u%
\mathcal{K}})^{\alpha }{}_{\mu }(e^{-u\mathcal{K}})^{\beta }{}_{\nu
}+(e^{(1-u)\mathcal{K}})^{\alpha }{}_{\mu }(e^{(1-u)\mathcal{K}})^{\beta
}{}_{\nu }\otimes M_{\alpha \beta }  \label{M-Delta} \\
& -\frac{\theta ^{\alpha \beta }}{2\kappa }(a_{\mu }\delta _{\nu }^{\gamma
}-a_{\nu }\delta _{\mu }^{\gamma })\left[ (1-u)p_{\delta }\otimes M_{\alpha
\beta }(e^{-u\mathcal{K}})^{\delta }{}_{\gamma }-uM_{\alpha \beta }(e^{(1-u)%
\mathcal{K}})^{\delta }{}_{\gamma }\otimes p_{\delta }\right] ,  \notag
\end{align}%
where $\mathcal{K}_{\mu \nu }$is given in equation \eqref{mathcalK}.

Corresponding antipodes \eqref{general-def-antip} 
are: 
\begin{align}
S_{\mathcal{F}}(p_{\mu })& =-{(e^{-(1-2u)\mathcal{K}})}^{\alpha }{}_{\mu
}p_{\alpha },  \label{S-p} \\
S_{\mathcal{F}}(M_{\mu \nu })& =-M_{\alpha \beta }{(e^{-(1-2u)\mathcal{K}%
})^{\alpha }{}_{\mu }(e^{-(1-2u)\mathcal{K}})}^{\beta }{}_{\nu }  \label{S-M}
\\
& +\frac{1}{\kappa }(a_{\mu }\delta _{\nu }^{\alpha }-a_{\nu }\delta _{\mu
}^{\alpha })\left[ S(p_{\alpha })+(1-2u)\theta _{\alpha \beta }S(p^{\beta })%
\right] ,  \notag
\end{align}

The counit is trivial: 
\begin{equation}
\epsilon (p_{\mu })=0,\qquad \epsilon (M_{\mu \nu })=0.
\end{equation}

It is interesting to note that coproduct and antipode of $\mathcal{K}_{\mu
\nu }$ remain classical, i.e. 
\begin{align}
\Delta _{\mathcal{F}}(\mathcal{K}_{\mu \nu })& =\mathcal{K}_{\nu }\otimes
1+1\otimes \mathcal{K}_{\mu \nu }=\Delta _{0}(\mathcal{K}_{\mu \nu }), \\
S_{\mathcal{F}}(\mathcal{K}_{\mu \nu })& =-\mathcal{K}_{\mu \nu }=S_{0}(%
\mathcal{K}_{\mu \nu }).
\end{align}%
Coproduct and antipode of $(e^{\mathcal{K}})_{\mu }{}^{\nu }$ are 
\begin{align}
\Delta _{\mathcal{F}}(e^{\mathcal{K}})_{\mu }{}^{\nu }& =(e^{\mathcal{K}%
})_{\mu }{}^{\alpha }\otimes (e^{\mathcal{K}})_{\alpha }{}^{\nu }, \\
S_{\mathcal{F}}((e^{\mathcal{K}})_{\mu }{}^{\nu })& =(e^{-\mathcal{K}})_{\mu
}{}^{\nu }.
\end{align}

\bigskip If noncommutativity is introduced through Hopf-algebraic twist
quantization one can introduce the star product realization of the algebra $%
\hat{A}$ of functions on quantum Minkowski space in terms of $\star $%
-algebra of classical functions $f(x),g(x)$%
\begin{equation}
f(x)\star _{\mathcal{F}}g(x)=m\left[ \mathcal{F}^{-1}(\triangleright \otimes
\triangleright )(f(x)\otimes g(x))\right] ,  \label{starprod}
\end{equation}%
where the action $\rhd $ is defined by eq. (\ref{act1}), (\ref{act3}) and $(%
\hat{A},\cdot )$ algebra is represented as $(A,\star )$ algebra.

Alternatively, one can write equation (\ref{starprod}) in the following form%
\begin{equation}
f(x)\star _{\mathcal{F}}g(x)=\hat{f}_{\mathcal{F}}\triangleright g(x),
\end{equation}%
where%
\begin{equation}
\hat{f}_{\mathcal{F}}=m\left[ \mathcal{F}^{-1}(\triangleright \otimes
1)(f(x)\otimes 1)\right] ,  \label{fdefor}
\end{equation}%
is a noncommutative counterpart of $f(x)$, described as the functions of
classical generators $(x_{\mu },p_{\nu },M_{\rho \sigma })\in \mathcal{P}\#%
\mathcal{M}$. Because $p_{\mu }\rhd 1=M_{\mu \nu }\rhd 1=0$ one gets that $%
\hat{x}_{\mu }\rhd 1=x_{\mu }$ and subsequently $\hat{f}_{\mathcal{F}}\rhd
1=f(x).$

Following (\ref{xTwist}), if we put in (\ref{fdefor}) $f=x_{\mu }$ and $%
\mathcal{F}=\mathcal{F}_{u}$ the non-commutative coordinates are given by%
\begin{equation}
\begin{split}
\hat{x}_{\mu }& =m\left[ \mathcal{F}_{u}^{-1}(\triangleright \otimes
1)(x_{\mu }\otimes 1)\right] \\
& =x_{\alpha }{(e^{-u\mathcal{K}})_{\mu }}^{\alpha }+(1-u)\frac{ia_{\mu }}{%
2\kappa }\theta ^{\alpha \beta }M_{\alpha \beta }.
\end{split}
\label{eqref}
\end{equation}%
The non-commutative coordinates (\ref{eqref}) close to a Lie algebra 
\begin{equation}
\lbrack \hat{x}_{\mu },\hat{x}_{\nu }]=\frac{i}{\kappa }(a_{\mu }\theta
_{\nu \alpha }-a_{\nu }\theta _{\mu \alpha })\hat{x}^{\alpha }.
\label{xx-def}
\end{equation}%
Note that the structure constants $C_{\mu \nu }{}^{\alpha }=\frac{1}{\kappa }%
(a_{\mu }\theta _{\nu }{}^{\alpha }-a_{\nu }\theta _{\mu }{}^{\alpha })$ do
not depend on the parameter $u$ and satisfy Jacobi identities.

If $u=1$ the twist $\mathcal{F}_{1}$ is special, because only such value of $%
u$ gives the particular choice described by formula (\ref{xdef})%
\begin{equation}
\hat{x}_{\mu }=x_{\alpha }(e^{-\mathcal{K}}{)_{\mu }}^{\alpha },
\end{equation}%
for any choice of the Lorentz generators $M_{\alpha \beta }$. If \ $u=0$,
eq. (\ref{eqref}) reduces to%
\begin{equation}
\hat{x}_{\mu }=x_{\mu }+\frac{ia_{\mu }}{2\kappa }\theta ^{\alpha \beta
}M_{\alpha \beta .}
\end{equation}%
This case was considered in \cite{kumar}\ and it was related to twisted
statistics. For $u=1/2$, twist $\mathcal{F}^{\tau }\mathcal{=F}^{-1}$ and
because $\mathcal{F}^{\dagger }=\mathcal{F}^{\tau }$ (where $^{\dagger }$
denotes Hermitian conjugation), for such a choice of $u$ the twist $\mathcal{%
F}_{\frac{1}{2}}$ is unitary. This case was considered in \cite{lw06},\cite%
{bala} . In \cite{bala} it was related to non-Pauli effects in
noncommutative spacetimes.

In order to obtain the coproduct sector and consistent bialgebroid relations
for $\hat{x}_{\mu }$ defined by (\ref{eqref}) we should look for the Hopf
algebroid structure of twist-deformed cross product $(\mathcal{P}\#\mathcal{M%
})_{\mathcal{F}}$.


\section{Deformed Heisenberg algebras and twisted cross products in
algebroid approach}

Quantum-mechanical phase-space coordinates $x^{\mu }$ and momenta $p_{\mu }$
describe canonical undeformed Heisenberg algebra, given by: 
\begin{equation}
\begin{split}
\lbrack x^{\mu },x^{\nu }]& =0, \\
\lbrack p_{\mu },x^{\nu }]& =-i\delta _{\mu }^{\nu }, \\
\lbrack p_{\mu },p_{\nu }]& =0.
\end{split}
\label{phase0}
\end{equation}%
If we deal with Hopf-algebraic scheme of Poincare symmetries the relations (%
\ref{phase0}) can be derived by the identification of $x_{\mu }\in \mathcal{M%
}$ with Abelian space-time translations of classical Poincare group and $%
p_{\mu }\in \mathcal{T}$ \ with the generators of \ dual Abelian fourmomenta
subalgebra acting on $\mathcal{M}$. The standard quantum-mechanical
phase-space with basic algebra (\ref{phase0}) is provided by smash product $%
\mathcal{H}_{0}\mathcal{=T}\#\mathcal{M}$, defining Heisenberg double with
undeformed (canonical) Heisenberg Hopf algebroid structure \footnote{%
It has been shown (see \cite{lu},Sect.6) that Heisenberg doubles are endowed
with Hopf algebroid structure.} of two Abelian dual Hopf algebras which
describe respectively the functions of coordinates $x_{\mu }$ and momenta $%
p_{\mu }$. The cross multiplication rules in $\mathcal{H}_{0}$ are given by
the Heisenberg double formula \cite{majid},\cite{luknow}%
\begin{equation}
p_{\mu }x_{\nu }=x_{\nu }^{(1)}\langle p_{\mu }^{(1)},x_{\nu }^{(2)}\rangle
p_{\mu }^{(2)},\qquad p_{\mu }\in \mathcal{T},\quad x_{\nu }\in \mathcal{M},
\label{HDform}
\end{equation}%
where $\langle \cdot ,\cdot \rangle $ describes the canonical duality
pairing, with $\Delta _{0}(p_{\mu })=p_{\mu }^{(1)}\otimes p_{\mu }^{(2)}$
given by (\ref{undef-antip}) and%
\begin{equation}
\Delta _{0}(x_{\mu })=x_{\mu }^{(1)}\otimes x_{\mu }^{(2)}=x_{\mu }\otimes
1+1\otimes x_{\mu }.  \label{zerokop}
\end{equation}%
The relation (\ref{zerokop}) follows as well from the coproduct of classical
Poincare group describing space-time translations, after contraction of the
Lorentz group parameters $\Lambda _{\nu }^{\mu }\longrightarrow \delta _{\nu
}^{\mu }$. The action $p_{\mu }\triangleright x_{\nu }$, given by formula (%
\ref{act3}), in Hopf-algebraic scheme can be identified with the binary
duality map $\mathcal{T\otimes M\longrightarrow \,}C:p\otimes x\rightarrow
\langle p,x\rangle $, which provides the differential realization of
fourmomenta $p_{\mu }$%
\begin{equation}
p_{\mu }\triangleright f(x)=\langle p_{\mu },f(x)\rangle =\frac{1}{i}%
\partial _{\mu }f(x).
\end{equation}%
Finally, one can easily deduce from (\ref{HDform}-\ref{zerokop}) and (\ref%
{act3}) the set of canonical commutation relations given by eq. (\ref{phase0}%
).

\bigskip In $\mathcal{H}_{0}$ one can choose different bases, in particular
\ it is possible to incorporate the change $x_{\mu }\longrightarrow \widehat{%
x}_{\mu }=x_{\rho }\varphi _{\ \mu }^{\rho }(p)$ (see (\ref{xdef})) and
employ the fourmomenta $p_{\mu }$, satisfying the relations (\ref{pkom})
with twisted coproduct (\ref{Delta-p}) of $p_{\mu }$ denoted as follows 
\begin{equation}
\Delta _{\mathcal{F}}(p_{\nu })\equiv \Delta _{\mathcal{F}}^{(1)}(p_{\nu
})\otimes \Delta _{\mathcal{F}}^{(2)}(p_{\nu })=\Delta _{0}(p_{\nu })+\delta
\Delta _{\mathcal{F}}(p_{\nu }).  \label{dec1}
\end{equation}%
As follows from (\ref{eqref}) the deformed Heisenberg algebra basis $(%
\widehat{x}^{\mu },p_{\mu })$ satisfies the standard duality relations $%
\langle p_{\mu },\widehat{x}_{\nu }\rangle =-i\eta _{\mu \nu }$. Further one
can show that the algebraic relation (\ref{xx-def}) are dual to the
coproducts (\ref{Delta-p}) in accordance with Hopf-algebraic duality, e.g.%
\begin{equation}
\langle \Delta _{\mathcal{F}}(p_{\rho }),\widehat{x}_{\mu }\otimes \widehat{x%
}_{\nu }\rangle =\langle p_{\rho },\widehat{x}_{\mu }\widehat{x}_{\nu
}\rangle .
\end{equation}%
Subsequently, introducing the coproduct%
\begin{equation}
\Delta (\widehat{x}_{\mu })=\widehat{x}_{\mu }^{(1)}\otimes \widehat{x}_{\mu
}^{(2)}=\widehat{x}_{\mu }\otimes 1+1\otimes \widehat{x}_{\mu },
\end{equation}%
which is dual to commuting fourmomenta $p_{\mu }$, one can show that we deal
with Heisenberg double $H$ with the basis $(\widehat{x}_{\mu },p_{\mu })$
describing deformed Heisenberg algebra. Therefore, the basic relation (\ref%
{HDform}) remains valid, i.e.%
\begin{equation}
p_{\mu }\widehat{x}_{\nu }=\widehat{x}_{\nu }^{(1)}\langle \Delta _{\mathcal{%
F}}^{(1)}(p_{\mu }),\widehat{x}_{\nu }^{(2)}\rangle \Delta _{\mathcal{F}%
}^{(2)}(p_{\mu }).  \label{HD22}
\end{equation}%
We get from decomposition (\ref{dec1}) that the term $\Delta _{0}(p_{\nu })$
gives the contribution $\eta _{\mu \nu }+\widehat{x}_{\nu }p_{\mu }$, and
relation (\ref{HD22}) takes the form of deformed canonical commutation
relations%
\begin{equation}
\lbrack p_{\mu },\widehat{x}_{\nu }]=-i\eta _{\mu \nu }+\left\{ (\Delta
^{(1)}(p)-\Delta _{(0)}^{(1)}(p))\vartriangleright \widehat{x}_{\nu
}\right\} (\Delta ^{(2)}(p)-\Delta _{(0)}^{(2)}(p)),  \label{phasede}
\end{equation}%
where due to the duality of coordinates $\widehat{x}_{\nu }$ and momenta $%
p_{\mu }$, we use the formula%
\begin{equation}
p_{\mu }\vartriangleright \widehat{x}_{\nu }=\langle p_{\mu },\widehat{x}%
_{\nu }\rangle =-i\eta _{\mu \nu }.
\end{equation}%
The formula (\ref{phasede}) can be also written in the form%
\begin{equation}
\lbrack p_{\mu },\widehat{x}_{\nu }]=-i\eta _{\mu \nu }+m[(\Delta -\Delta
_{0})(p_{\mu })(\vartriangleright \otimes 1)(\widehat{x}_{\nu }\otimes 1)],
\label{phasede2}
\end{equation}%
which was derived in alternative way also in \cite{mss}.

If we use the formulae (\ref{eqref}), (\ref{act2}) and (\ref{phase0}) we can
directly calculate the commutator (\ref{phasede2}). Such a method leads to
the same deformed Heisenberg algebra $\hat{\mathcal{H}}$, given by relations
(\ref{xx-def}), i.e. the cross commutator $[p_{\mu },\widehat{x}_{\nu }]$
which for any $u$ does not depend on Lorentz generators $M_{\mu \nu }$. We
obtain%
\begin{eqnarray}
\lbrack p_{\mu },\hat{x}_{\nu }] &=&-i(e^{-u\mathcal{K}})_{\nu \mu }+(1-u)%
\frac{ia_{\nu }}{\kappa }{\theta _{\mu }}^{\alpha }p_{\alpha },
\label{defo_ps} \\
\lbrack p_{\mu },p_{\nu }] &=&0.  \notag
\end{eqnarray}%
Additionally, commutation relations between $(e^{\mathcal{K}})_{\mu }{}^{\nu
}$ and $\hat{x}^{\lambda }$ and $M_{\alpha \beta }$ after using (\ref{act1}%
)-(\ref{act3}) are given by 
\begin{align}
\lbrack (e^{\mathcal{K}})_{\mu }{}^{\nu },\hat{x}^{\lambda }]& =-\frac{%
ia^{\lambda }}{\kappa }\theta _{\mu }{}^{\alpha }(e^{\mathcal{K}})_{\alpha
}{}^{\nu }, \\
\lbrack (e^{\mathcal{K}})_{\mu }{}^{\nu },M_{\alpha \beta }]& =\frac{%
a_{\alpha }p_{\beta }-a_{\beta }p_{\alpha }}{\kappa }(\theta e^{\mathcal{K}%
})_{\mu }{}^{\nu },
\end{align}%
and the commutation relations between Lorentz generators $M_{\mu \nu }$ and
non-commutative coordinates $\hat{x}_{\rho }$ are the following form 
\begin{equation}
\begin{split}
\lbrack M_{\mu \nu },\hat{x}_{\rho }]& =(\delta _{\mu }^{\alpha }\eta
_{\beta \nu }-\delta _{\nu }^{\alpha }\eta _{\beta \mu })\left[ \hat{x}%
_{\gamma }{(e^{u\mathcal{K}})_{\alpha }}^{\gamma }-(1-u)\frac{ia_{\alpha }}{%
2\kappa }\theta ^{\gamma \delta }M_{\gamma \delta }\right] {(e^{-u\mathcal{K}%
})_{\rho }}^{\beta } \\
& ~~+\frac{1}{\kappa }u{\theta _{\rho }}^{\alpha }\hat{x}_{\alpha }(a_{\mu
}p_{\nu }-a_{\nu }p_{\mu })+(1-u)ia_{\rho }({M_{\mu }}^{\alpha }\theta
_{\alpha \nu }-{\theta _{\mu }}^{\alpha }M_{\alpha \nu }).
\end{split}
\label{mxx}
\end{equation}%
The relation (\ref{mxx}) in generalized quantum-deformed phase space $(\hat{x%
}_{\mu },p_{\mu },S_{\mu \nu })$ where%
\begin{equation}
M_{\mu \nu }=i(x_{\mu }p_{\nu }-x_{\nu }p_{\mu })+S_{\mu \nu },
\end{equation}%
describe the noncommutativity of translational and spin degrees of freedom.

Let us consider now the Hopf algebroid $\mathcal{H}_{\mathcal{F}}$\ with
algebraic structure described by classical Poincare algebra $\mathcal{P}$\
supplemented by the noncommutative space-time coordinates $\hat{x}_{\mu }\in 
\mathcal{\hat{M}}$ satisfying the relations (\ref{xx-def}), (\ref{defo_ps})
and (\ref{mxx})%
\begin{equation}
\mathcal{H}_{\mathcal{F}}=(\mathcal{A},m;\mathcal{B}_{\mathcal{F}},s_{%
\mathcal{F}},t_{\mathcal{F}},\tilde{\Delta}_{\mathcal{F}},\tilde{\epsilon}_{%
\mathcal{F}},S_{\mathcal{F}}).  \label{hopfalg}
\end{equation}%
The total algebra $\mathcal{A}$ with the basis $(\hat{x}_{\mu },p_{\mu
},M_{\mu \nu })$ is given by the smash product $\mathcal{U}(\mathcal{P)}\#%
\mathcal{U}(\mathcal{\hat{M}})$ and base algebra $\mathcal{B}_{\mathcal{F}}$ 
$(\hat{x}_{\mu }\in \mathcal{B}_{\mathcal{F}})$ is provided by the algebra
of functions on $\mathcal{\hat{M}}$ with the multiplication in $\mathcal{B}_{%
\mathcal{F}}$ represented by star product formula (\ref{starprod}). The
source map $s_{\mathcal{F}}:\mathcal{B}_{\mathcal{F}}\mathcal{\rightarrow A}$
(algebra homomorphism) \ and target map $t_{\mathcal{F}}:\mathcal{B}_{%
\mathcal{F}}\mathcal{\rightarrow A}$ (algebra antihomomorphism) introduce in 
$\mathcal{A}$ the $(\mathcal{B}_{\mathcal{F}},\mathcal{B}_{\mathcal{F}})$
bimodule structure, namely for any $a\in \mathcal{A}$ and $b,b^{\prime }\in 
\mathcal{B}_{\mathcal{F}}$ one gets the formula $bab^{\prime }=s_{\mathcal{F}%
}(b)t_{\mathcal{F}}(b^{\prime })a$, i.e. we consider $\mathcal{H}_{\mathcal{F%
}}$ as left bialgebroid \cite{bohm},\cite{lu}.The comultiplication map $%
\tilde{\Delta}_{\mathcal{F}}:\mathcal{A\rightarrow A\otimes }_{\mathcal{B}_{%
\mathcal{F}}}\mathcal{A}$ with nonstandard tensor product introduced firstly
in \cite{takeu}\ is a coassociative bimodule map with the elements $a%
\mathcal{\otimes }_{\mathcal{B}_{\mathcal{F}}}a^{\prime }\in \mathcal{%
A\otimes }_{\mathcal{B}_{\mathcal{F}}}\mathcal{A}$ defined in the
description using standard tensor product $\mathcal{A\otimes A}$ by the
equivalence class generated by the following condition \cite{lu}%
\begin{equation}
m(\mathcal{I}_{\mathcal{F}}(b\mathcal{\otimes }b^{\prime }))=0,\qquad 
\mathcal{I}_{\mathcal{F}}=\mathcal{(}t_{\mathcal{F}}\otimes 1-1\otimes s_{%
\mathcal{F}}),  \label{takpr}
\end{equation}%
where $\mathcal{A\otimes }_{\mathcal{B}_{\mathcal{F}}}\mathcal{A=(A\otimes
A)\diagup I}_{\mathcal{F}}$ and $s_{\mathcal{F}}$ and $t_{\mathcal{F}}$ are
respectively the twisted source and target maps defined below (see (\ref%
{sour})-(\ref{targ})). If we describe coproducts\footnote{%
We denote the bialgebroid coproducts with tilda.} $\tilde{\Delta}_{\mathcal{F%
}}$ using standard tensor products one can treat the elements $(a\mathcal{%
\otimes }a^{\prime })$ satisfying (\ref{takpr}) as defining coproduct
gauges, with gauge-invariant elements described by $a\mathcal{\otimes }_{%
\mathcal{B}_{\mathcal{F}}}a^{\prime }$. In particular for $\hat{x}_{\mu }\in 
\mathcal{B}_{\mathcal{F}}$ we shall choose the special coproduct gauge
defined by the formula (see e.g. \cite{lu},\cite{lws})%
\begin{equation}
\tilde{\Delta}_{\mathcal{F}}(\widehat{x}_{\mu })=\widehat{x}_{\mu }\otimes 1.
\label{bix}
\end{equation}%
The canonical choice of the coproduct given by the formula (\ref{bix}) can
be obtained if we insert the twisted coproducts (\ref{Delta-p})-(\ref%
{M-Delta}) and%
\begin{equation}
\tilde{\Delta}_{\mathcal{F}}(x_{\mu })=\mathcal{F}\tilde{\Delta}_{0}(x_{\mu
})\mathcal{F}^{-1},\qquad \tilde{\Delta}_{0}(x_{\mu })=x_{\mu }\otimes 1,
\end{equation}%
into the relation (\ref{eqref}), in accordance with the equality%
\begin{equation}
\widehat{x}_{\mu }\equiv \hat{x}_{\mu }(x_{\mu },p_{\mu },M_{\mu \nu
})\longrightarrow \tilde{\Delta}_{\mathcal{F}}(\widehat{x}_{\mu })\equiv 
\hat{x}_{\mu }(\tilde{\Delta}_{\mathcal{F}}(x_{\mu }),\Delta _{\mathcal{F}%
}(p_{\mu }),\Delta _{\mathcal{F}}(M_{\mu \nu })).
\end{equation}%
One can check further that the coproducts (\ref{bix}) and (\ref{Delta-p})-(%
\ref{M-Delta}) describe the homomorphic map $\mathcal{A\rightarrow A\otimes A%
}$ of the algebraic relations (\ref{xx-def}), (\ref{defo_ps}) and (\ref{mxx}%
) i.e. if $a=(\hat{x}_{\mu },p_{\mu },M_{\mu \nu })\in \mathcal{A}$ we get
the following canonical full set of coproducts for bialgebroid (\ref{hopfalg}%
) (see (\ref{bix}) and (\ref{Delta-p})-(\ref{M-Delta}))%
\begin{equation}
\tilde{\Delta}_{\mathcal{F}}(a)=(\tilde{\Delta}_{\mathcal{F}}(\widehat{x}%
_{\mu }),\tilde{\Delta}_{\mathcal{F}}(p_{\mu })=\Delta _{\mathcal{F}}(p_{\mu
}),\tilde{\Delta}_{\mathcal{F}}(M_{\mu \nu })=\Delta _{\mathcal{F}}(M_{\mu
\nu })).  \label{allcop}
\end{equation}%
The twisted source and target maps are introduced as follows \cite{lu},\cite%
{xu},\cite{bp}%
\begin{eqnarray}
s_{0}(x_{\mu })=x_{\mu }\qquad \overset{\mathcal{F}}{\rightarrow }\qquad s_{%
\mathcal{F}}(\widehat{x}_{\mu }) &=&m[\mathcal{F}^{-1}(\vartriangleright
\otimes 1)(s_{0}(x_{\mu })\otimes 1)]=\widehat{x}_{\mu },  \label{sour} \\
t_{0}(x_{\mu })=x_{\mu }\qquad \overset{\mathcal{F}}{\rightarrow }\qquad t_{%
\mathcal{F}}(\widehat{x}_{\mu }) &=&m[(\mathcal{F}^{-1})^{\tau
}(\vartriangleright \otimes 1)(t_{0}(x_{\mu })\otimes 1)]  \label{targ} \\
&=&\widehat{x}_{\alpha }{(e^{\mathcal{K}})_{\mu }}^{\alpha }-i\frac{a_{\mu }%
}{2\kappa }\theta ^{\alpha \beta }M_{\alpha \beta }.  \notag
\end{eqnarray}%
Due to the model-independent relation (see e.g. \cite{lu}, proof of
preposition $(2.4)$)%
\begin{equation}
\tilde{\Delta}_{\mathcal{F}}(s_{\mathcal{F}}(\widehat{x}_{\mu }))=s_{%
\mathcal{F}}(\widehat{x}_{\mu })\otimes 1,
\end{equation}%
it follows that the formulae (\ref{sour}) and (\ref{bix}) are consistent as
expected. Further, it can be shown that the source and target maps satisfy
the relations $(C{_{\mu \nu }}^{\alpha }=\frac{i}{\kappa }a_{[\mu }\theta {%
_{\nu ]}}^{\alpha })$%
\begin{eqnarray}
\lbrack s(\widehat{x}_{\mu }),s(\widehat{x}_{\nu })] &=&C{_{\mu \nu }}%
^{\alpha }s(\widehat{x}_{\alpha }), \\
\lbrack t(\widehat{x}_{\mu }),t(\widehat{x}_{\nu })] &=&-C{_{\mu \nu }}%
^{\alpha }t(\widehat{x}_{\alpha }), \\
\lbrack s(\widehat{x}_{\mu }),t(\widehat{x}_{\nu })] &=&0.
\end{eqnarray}%
\qquad The coproduct $\tilde{\Delta}_{\mathcal{F}}:\mathcal{A\rightarrow
A\otimes A}$ is only coassociative when $\mathcal{A\otimes A}$ is projected
into equivalence classes $\mathcal{A\otimes }_{\mathcal{B}_{\mathcal{F}}}%
\mathcal{A}$ generated by the ideal $\mathcal{I}_{\mathcal{F}}$. The choice
of representatives in the equivalence class defines the coproduct gauge.

The simplest choice of the coproduct gauge transformation is obtained by
adding to (\ref{bix}) the ideal $\mathcal{I}_{\mathcal{F}}$ multiplied by a
constant $\alpha $ 
\begin{equation}
\tilde{\Delta}_{(\alpha )}(\widehat{x}_{\mu })=\widehat{x}_{\mu }\otimes
1+\alpha (t_{\mathcal{F}}(\widehat{x}_{\mu })\otimes 1-1\otimes s_{\mathcal{F%
}}(\widehat{x}_{\mu })).
\end{equation}

One can check that the coproducts $\tilde{\Delta}_{(\alpha )}(\widehat{x}%
_{\mu })$ together with the Hopf-algebraic coproducts (\ref{Delta-p})-(\ref%
{M-Delta}) satisfy the algebraic relations which are homomorphic to the
relations (\ref{xx-def}), (\ref{defo_ps}) and (\ref{mxx}). The coproduct
gauge can be generalized by introducing powers of ideal $\mathcal{I}_{%
\mathcal{F}}$ as well as powers of coproducts (\ref{allcop}). In such a case
the homomorphism between the algebraic and coalgebraic relations of Hopf
algebroid $\mathcal{H}_{\mathcal{F}}$ will be only valid in standard tensor
notation modulo the choices of coproduct gauge transformation \cite{jkm},%
\cite{lws}.

Finally we complete the description of Hopf bialgebroid $\mathcal{H}_{%
\mathcal{F}}$ structure if we define ($\epsilon _{0}(x_{\mu })=x_{\mu }$ \
for undeformed case) (see e.g. \cite{jkm}) 
\begin{eqnarray}
\epsilon _{\mathcal{F}}(\widehat{x}_{\mu }) &=&m[\mathcal{F}%
^{-1}(\triangleright \otimes 1)(\epsilon _{0}(x_{\mu })\otimes 1)]=\widehat{x%
}_{\mu },  \label{jed} \\
\epsilon _{\mathcal{F}}(p_{\mu }) &=&\epsilon _{\mathcal{F}}(M_{\mu \nu
})=0,\qquad \epsilon _{\mathcal{F}}(1)=1.
\end{eqnarray}%
Hopf algebroid is a bialgebroid with antipode (coinverse). In order to
obtain the antipode $S_{\mathcal{F}}$\ one can use the formula $(a\in 
\mathcal{A)}$ (see (\ref{general-def-antip}))%
\begin{equation}
S_{\mathcal{F}}(a)=\chi _{\mathcal{F}}S_{0}(a)\chi _{\mathcal{F}}^{-1},
\label{antieq}
\end{equation}%
with $\chi _{\mathcal{F}}=\exp [-(1-2u)\frac{a\cdot p}{2\kappa }\theta
^{\alpha \beta }M_{\alpha \beta }]$, one gets%
\begin{equation}
S_{\mathcal{F}}(\widehat{x}_{\mu })={(e^{\mathcal{K}})_{\mu }}^{\alpha }%
\widehat{x}_{\alpha }-i\frac{a_{\mu }}{2\kappa }\theta ^{\alpha \beta
}M_{\alpha \beta }=t_{\mathcal{F}}(\widehat{x}_{\mu }).
\end{equation}%
Note that $S_{\mathcal{F}}^{2}=1$. The antipodes $S_{\mathcal{F}}(p_{\mu })$%
, $S_{\mathcal{F}}(M_{\mu \nu })$ in Hopf algebroid (\ref{hopfalg}) remain
the same as for the twisted Poincare -Hopf algebra (see (\ref{S-p})-(\ref%
{S-M})) and are also involutive (see (\ref{general-def-antip})). Further it
can be shown that 
\begin{eqnarray}
S_{\mathcal{F}}(t_{\mathcal{F}}(\widehat{x}_{\mu })) &=&s_{\mathcal{F}}(%
\widehat{x}_{\mu })=\widehat{x}_{\mu },  \label{ant1} \\
m[(1\otimes S_{\mathcal{F}})\circ \tilde{\Delta}_{\mathcal{F}}] &=&s_{%
\mathcal{F}}\epsilon _{\mathcal{F}}=\epsilon _{\mathcal{F}},  \label{antt} \\
m[(S_{\mathcal{F}}\otimes 1)\circ \tilde{\Delta}_{\mathcal{F}}] &=&t_{%
\mathcal{F}}\epsilon _{\mathcal{F}}S_{\mathcal{F}}.  \label{ant2}
\end{eqnarray}%
Note that in second formula the introduction of anchor projection $\gamma $ (%
$\tilde{\Delta}_{\mathcal{F}}\rightarrow \gamma \tilde{\Delta}_{\mathcal{F}} 
$, \ where $\gamma $ is a section of the projection $\mathcal{A\otimes
A\rightarrow A\otimes }_{\mathcal{B}_{\mathcal{F}}}\mathcal{A},$ see \cite%
{lu}-\cite{brzez}) is not needed\footnote{%
The anchor projection restricts the coproduct gauge for which the formula (%
\ref{antt}) is valid.}.

Finally we recall that for spinless systems one can introduce the orbital
realization of Lorentz generators $M_{\mu \nu }$, described by formula (\ref%
{repM}). Inserting (\ref{repM}) in formula (\ref{twist-F}) one gets the $u$%
-dependent twist of canonical Heisenberg-Hopf algebroid $\mathcal{H}_{%
\mathcal{F}}$%
\begin{equation}
\mathcal{F}_{u}(p_{\mu },M_{\mu \nu })\rightarrow \widetilde{\mathcal{F}}%
_{u}(p_{\mu },x_{\nu })=\mathcal{F}_{u}(p_{\mu },i(x_{\mu }p_{\nu }-x_{\nu
}p_{\mu })),  \label{xptwist}
\end{equation}%
with the coproducts $\tilde{\Delta}_{\mathcal{\tilde{F}}}$ of $x_{\mu }$ and 
$p_{\nu }$ obtained from (\ref{Delta-p}) and (\ref{bix}) after inserting the
substitution (\ref{xptwist}). Further, it follows that the two-cocycle
condition of $\mathcal{F}$ (see \cite{majid}) is becoming a two-cocycle
condition for bialgebroid twist $\widetilde{\mathcal{F}}$ (see $\footnote{%
The two-cocycle condition for bialgebroid twists is given e.g. in \cite{xu},%
\cite{bp}.}$), given by (\ref{xptwist}).

It is easy to see that after inserting (\ref{repM}) in relation (\ref{eqref}%
) the formula (\ref{xdef}) becomes valid for all values of $u$. Concluding,
from the Hopf algebroid (\ref{hopfalg}) with independent Lorentz generators
by using (\ref{repM}) one obtains twisted Heisenberg-Hopf algebroid with the
formulae for source and target maps, antipodes and the ideal describing
coproduct gauges expressed only in terms of phase space variables $(\hat{x}%
_{\mu },p_{\mu })$ or $(x_{\mu },p_{\mu })$ $\footnote{%
The canonical coordinates $x_{\mu }$ can be expressed in terms of $\widehat{x%
}_{\mu }$ if the formula (\ref{xdef}) is invertible.}$.

\section{Final Remarks}

The cross product algebra $\mathcal{P}\#\mathcal{M}$, with the algebra basis
described by generators $(p_{\mu },M_{\mu \nu },\hat{x}_{\mu })$, can be
named Poincare-Heisenberg algebra \cite{lunow2} or DSR algebra \cite{gkn},%
\cite{bp2}\footnote{%
DSR $\equiv $ Doubly Special Relativity or Deformed Special Relativity.}. In
this paper we provide a particular example of quantum twist-deformed DSR
algebra $\mathcal{P}\#\mathcal{\hat{M}}$ and present explicitly its
algebraic and coalgebraic Hopf algebroid structure. It should be observed
that DSR\ algebra can be obtained by the contraction of full generalized
relativistic quantum phase space described as the Heisenberg double (see
e.g. \cite{lws}), i.e. the cross product $\mathbb{H}\#\mathbb{\tilde{H}}$\
of deformed Poincare-Hopf algebra $\mathbb{H}$ (with basis $(p_{\mu },M_{\mu
\nu })$) and quantum Poincare-Hopf quantum group $\mathbb{\tilde{H}}$ (with
basis $(x_{\mu },\Lambda _{\mu \nu })$, where $\Lambda _{\mu \nu }$ ($%
\Lambda _{\mu \alpha }\Lambda _{\;\nu }^{\alpha }=\eta _{\mu \nu }$)
describe the Lorentz $4\times 4$ matrix group elements).

Because $D=4$ Heisenberg algebra is described as well by the cross-product $%
\mathcal{T}_{4}\#\mathcal{M}_{4}$, one can represent $D=4$ DSR algebraic
structure as the following composition of cross products%
\begin{equation}
DSR\;algebra=SO(3,1)\#(\mathcal{T}_{4}\#\mathcal{M}_{4}).  \label{dsr}
\end{equation}%
It follows from (\ref{dsr}) that the Lorentz generators $SO(3,1)$ act
covariantly on the standard (without spin degrees of freedom) quantum phase
space $\mathcal{T}_{4}\#\mathcal{M}_{4}$. We add that the cross-product
structures presented in (\ref{dsr}) are preserved for twist quantum-deformed
phase space $(\mathcal{T}_{4}\#\mathcal{\hat{M}}_{4})_{\mathcal{F}}$ endowed
with quantum-relativistic covariance under the action of twisted Poincare
-Hopf algebra

There remain some questions which should be further studied. In particular
one should elaborate more on the role of Heisenberg algebra twists (see e.g.
(\ref{xptwist})), in the construction of Hopf algebroids which provide the
quantum deformed relativistic phase space frameworks. In this paper the
advantage of our approach to quantum phase space formulation is the
appearance of spin degrees of freedom $S_{\mu \nu }$ as independent phase
space coordinates. The extension of quantum phase spaces with spin degrees
of freedom still remains quite open subject, and we plan to study the
relation of such extended phase spaces (see e.g. \cite{souri},\cite{bette}
in undeformed case) with the Hopf algebroid constructions.

\section*{Acknowledgements}

One of the authors (J.L.) would like to thank Andrzej Borowiec for valuable
comments. J.L and M.W. have been suported by Polish National Science Center,
project 2014/13/B/ST2/04043 and the work by S.M. and D.P. has been supported
by Croatian Science Foundation under the Project No. IP-2014-09-9582 as well
as by the H2020 Twinning project No. 692194, \textquotedblleft
RBI-T-WINNING\textquotedblright .

\end{document}